\documentclass[aps,prl,twocolumn,nofootinbib,showpacs,superscriptaddress]{revtex4-1}
\usepackage{graphicx}
\usepackage{amsmath}

\begin{document}

\title{A search for varying fundamental constants using Hz-level frequency measurements of cold CH molecules}
\author{S. Truppe}
\affiliation{Centre for Cold Matter, Blackett Laboratory, Imperial College London, Prince Consort Road, London, SW7 2AZ, United Kingdom.}
\author{R. J. Hendricks}
\affiliation{Centre for Cold Matter, Blackett Laboratory, Imperial College London, Prince Consort Road, London, SW7 2AZ, United Kingdom.}
\author{S. K. Tokunaga}
\affiliation{Centre for Cold Matter, Blackett Laboratory, Imperial College London, Prince Consort Road, London, SW7 2AZ, United Kingdom.}
\author{H. J. Lewandowski}
\affiliation{JILA and Department of Physics, University of Colorado, 440 UCB, Boulder, CO 80309-0440, United States.}
\author{M. G. Kozlov}
\affiliation{Petersburg Nuclear Physics Institute, Gatchina 188300, Russia.}
\author{Christian Henkel}
\affiliation{Max-Planck-Institut f{\"u}r Radioastronomie, Auf dem H{\"u}gel 69,
53121 Bonn, Germany.}
\affiliation{ Astronomy Department, King Abdulaziz University, P.O. Box 80203, Jeddah, Saudi Arabia. }
\author{E. A. Hinds}
\affiliation{Centre for Cold Matter, Blackett Laboratory, Imperial College London, Prince Consort Road, London, SW7 2AZ, United Kingdom.}
\author{M. R. Tarbutt}\email{m.tarbutt@imperial.ac.uk}
\affiliation{Centre for Cold Matter, Blackett Laboratory, Imperial College London, Prince Consort Road, London, SW7 2AZ, United Kingdom.}

\begin{abstract}
Many modern theories predict that the fundamental constants depend on time, position, or the local density of matter. We develop a spectroscopic method for pulsed beams of cold molecules, and use it to measure the frequencies of microwave transitions in CH with accuracy down to $3\,$Hz. By comparing these frequencies with those measured from sources of CH in the Milky Way, we test the hypothesis that fundamental constants may differ between the high and low density environments of the Earth and the interstellar medium. For the fine structure constant we find $\Delta\alpha/\alpha = (0.3 \pm 1.1) \times 10^{-7}$, the strongest limit to date on such a variation of $\alpha$. For the electron-to-proton mass ratio we find $\Delta\mu/\mu = (-0.7 \pm 2.2) \times 10^{-7}$. We suggest how dedicated astrophysical measurements can improve these constraints further and can also constrain temporal variation of the constants.
\end{abstract}


\maketitle

In the standard model of particle physics, the fundamental constants are fixed parameters, but in many extensions of the standard model, they may change \cite{Uzan(1)03, Uzan(1)11}. In higher dimensional theories for example, which aim at a unified description of gravity and the other forces, the constants depend on the size of the compactified extra dimensions. Their size may be changing, in a conceptually similar way to the observed expansion of the Universe, causing the constants to change with position or time. Some theories of dark energy also predict variation of the constants. These theories hypothesize that a scalar field is responsible for the observed cosmic acceleration. In chameleon models, this field acquires an effective mass that depends strongly on the local matter density \cite{Khoury(1)04, Brax(1)04}. In our high density environment, the field has such short range that it goes undetected in tests of the equivalence principle and searches for fifth forces. In these models, particle masses and coupling constants may also depend on matter density \cite{Olive(1)08}. Given the potential role of these theories in answering outstanding questions in physics, it is important to test the prediction of varying constants.

Atomic and molecular transition frequencies provide a natural way to look for variation of the fine structure constant, $\alpha$, and the electron-proton mass ratio, $\mu$. Their present-day time derivatives are strongly constrained by comparisons of  atomic clocks \cite{Rosenband(1)08, Blatt(1)08}, while observations of high red-shift objects test changes on cosmological scales. Some analyses of optical absorption spectra at high red-shifts suggest a variation of $\alpha$ across the Universe \cite{Webb(1)11}, but others find no variation at the level of a few parts-per-million (ppm) \cite{Molaro(1)08, Agafonova(1)11}. High red-shift measurements using microwave and mm-wave molecular transitions, which can be far more sensitive to changes in the constants than optical transitions, give null results for $\Delta\alpha/\alpha$ at levels between 8 and 0.8\,ppm \cite{Levshakov(1)12, Weiss(1)12, Kanekar(1)10, Rahmani(1)12}. For $\Delta \mu/\mu$, a recent study of methanol absorption lines at high red-shift found no variation at the 0.1\,ppm level \cite{Bagdonaite(1)13}.

Measurements within our galaxy test whether constants depend on the local density. For changes in $\alpha$ with matter density the $1\sigma$ limit is $|\Delta \alpha/\alpha| < 2 \times 10^{-7}$ \cite{Levshakov(3)10}. For changes in $\mu$ a study of methanol maser lines sets a $1\sigma$ upper limit $|\Delta \mu/\mu|<2.9 \times 10^{-8}$ \cite{Levshakov(1)11}. However, Levshakov et al. \cite{Levshakov(1)10, Levshakov(2)10}, comparing terrestrial and astrophysical microwave transitions in ammonia and other molecules, find an eight-standard-deviation systematic difference. This suggests a fractional change in $\mu$ of $2.6 \times 10^{-8}$ when going from the earth to the interstellar medium, tentatively supporting the chameleon hypothesis.  This intriguing situation calls out for new, high-precision measurements of the most sensitive molecular transition frequencies and for higher quality astronomical observations.

Here, we measure the frequencies of the 3.3 and 0.7\,GHz $\Lambda$-doublet transitions of CH, which are very sensitive to changes in $\alpha$ and $\mu$ \cite{Kozlov(1)09, Nijs(1)12}. With these measurements, and existing astronomical observations of CH in the interstellar medium, we constrain variation of the constants in our Galaxy, and open the way for future radio-astronomy measurement of higher accuracy to test more stringently whether the constants vary. Our spectroscopic method has Hz-level accuracy, and takes advantage of the short pulse duration and small velocity spread of cold molecular beams to control the interaction between molecules and an applied microwave field. This method can also be applied to other molecules of interest in chemical physics, metrology, particle physics and cosmology \cite{Shelkovnikov(1)08, Bethlem(1)08, Bethlem(1)09, Dickenson(1)13, Salumbides(1)13}.

\begin{figure}
\includegraphics[width=\linewidth]{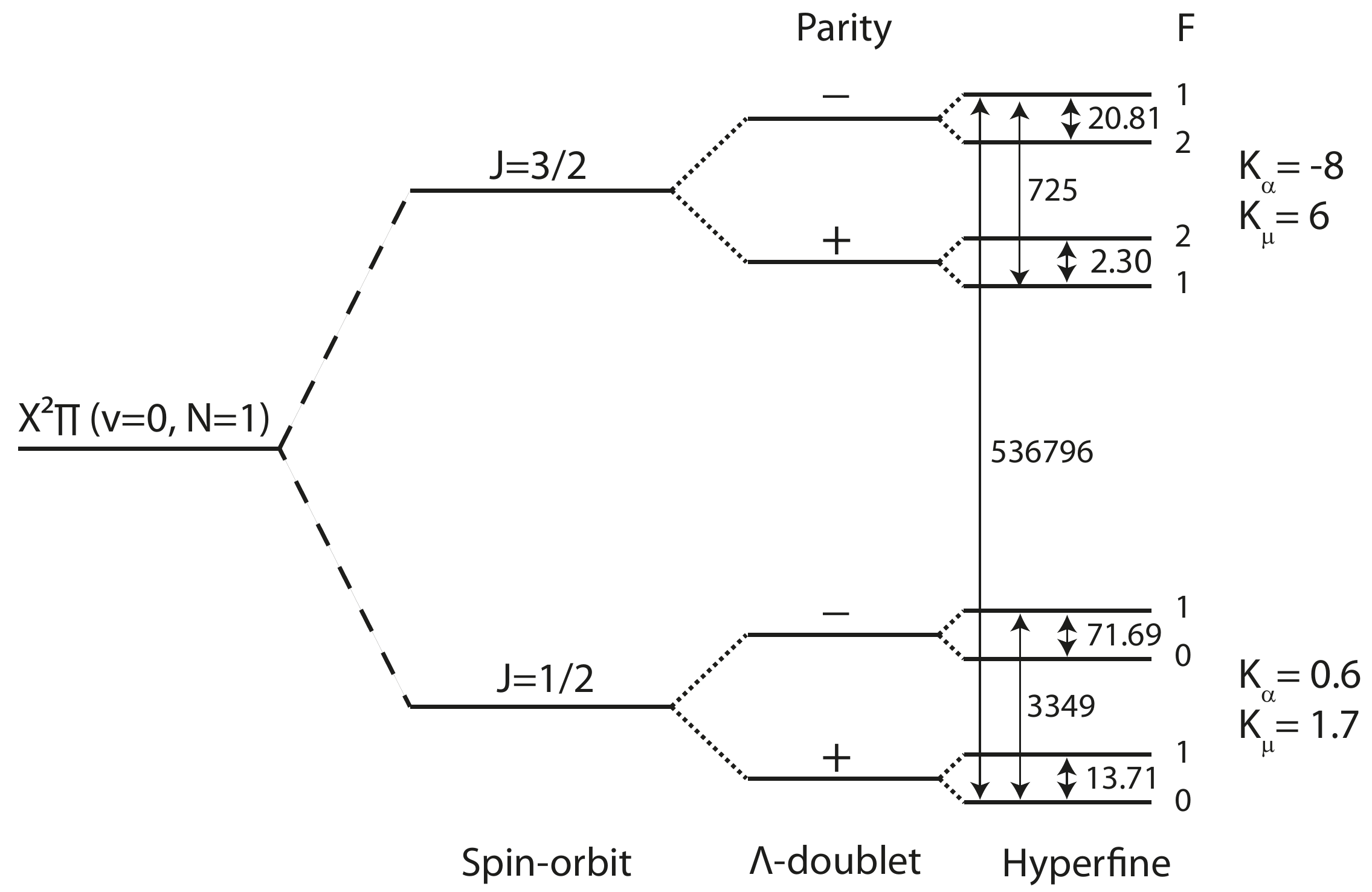}
\caption{\label{Fig:Levels} Relevant energy levels in CH. Approximate frequencies are given in MHz and the sensitivity coefficients for the two $\Lambda$-doublet transitions are shown.}
\end{figure}

\textbf{The CH molecule.} Figure \ref{Fig:Levels} shows the relevant energy levels in the ground electronic and vibrational state of CH, $\text{X}^{2}\Pi (v=0)$. These eigenstates of the Hamiltonian are also eigenstates of the parity operator, $P$, and of the squared total angular momentum operator, ${\bf F}^{2}$. In addition, they are approximate eigenstates of ${\bf N}^{2}$ and ${\bf J}^{2}$ where $\bf{J}=\bf{N}+\bf{S}$ and $\bf{N} = \bf{L} + \bf{R}$. Here, $\bf{S}$ and $\bf{L}$ are the electronic spin and orbital angular momentum operators, and $\bf{R}$ is the rotational angular momentum operator. The spin-orbit interaction splits the $N=1$ level into a pair of levels labelled by $J=1/2$ and $J=3/2$. Each is split by a Coriolis interaction into a $\Lambda$-doublet whose components are the parity eigenstates, $|p\!=\!\pm 1\rangle = |\Lambda\!=\!+1\rangle\!\pm\!(-1)^{J-S}|\Lambda\!=\!-1\rangle$, where $\Lambda$ is the quantum number for the projection of $\bf{L}$ onto the internuclear axis. It is these $\Lambda$-doublet frequencies that we measure. The interaction with the hydrogen nuclear spin, $I=1/2$, produces hyperfine structure in each of the $\Lambda$-doublet components. The levels are labelled as $(J^p,F)$.

These two $\Lambda$-doublet transitions have been observed towards numerous clouds in the interstellar medium \cite{Rydbeck(1)73, Turner(1)74, Robinson(1)74, Rydbeck(1)74, Zuckerman(1)75, Rydbeck(1)76, Hjalmarson(1)77, Whiteoak(1)78, Genzel(1)79, Ziurys(1)85}, and the 3.3\,GHz transition has also been observed in other galaxies \cite{Whiteoak(1)80, Nieves(1)13}. Figure \ref{Fig:Levels} gives the sensitivity coefficients of the transitions to changes in the constants, $K_{\alpha}$ and $K_{\mu}$ \cite{Kozlov(1)09, Nijs(1)12}, defined such that the rest frequency emitted by molecules in the astronomical source is $\omega_{\text{lab}}(1+ K_{\alpha} \Delta\alpha/\alpha + K_{\mu} \Delta \mu/\mu)$, where $\omega_{\text{lab}}$ is the laboratory frequency. Since these coefficients are large and different, a very sensitive test is possible when the transitions are observed together in the same gas. Until now, this test was hindered by a lack of precise laboratory frequencies.

\begin{figure*}
\includegraphics[width=\linewidth]{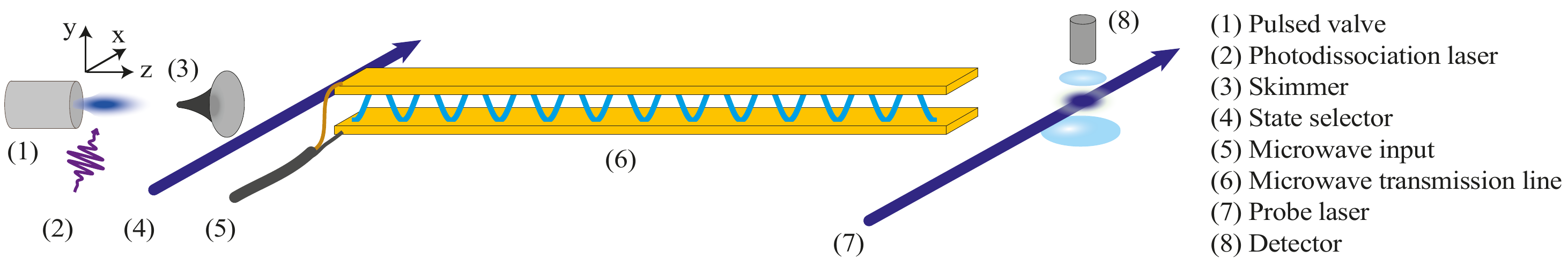}
\caption{\label{Fig:Setup} Experimental setup for measuring the $\Lambda$-doublet frequencies. A supersonic CH beam is produced by photo-dissociation of CHBr$_3$. The initial state is selected either by optical pumping (for $J=1/2$) or by driving the $(1/2^-)$--$(3/2^+)$ transition with a mm-wave beam (for $J=3/2$). The molecules travel through a microwave transmission line where the $\Lambda$-doublet transitions are driven, and are finally detected by laser induced fluorescence. Not to scale.}
\end{figure*}

\textbf{Frequency measurements.} Using the apparatus shown in Figure \ref{Fig:Setup}, we measure the laboratory frequencies by the Ramsey method of separated oscillating fields \cite{Ramsey(1)49, Ramsey(1)50}. The source produces short, cold pulses of CH molecules, which pass through a state selector, a transmission line resonator, and then a detector. The state selector preferentially populates one of the two parity eigenstates of the $\Lambda$-doublet being measured. Inside the transmission line resonator the molecules interact with a standing-wave microwave field. A first microwave pulse (a $\pi/2$ pulse), of duration $\tau=15$\,$\mu$s and angular frequency $\omega$, is applied when the molecules are at antinode $m_1$ of the field. This prepares an equal superposition of the two $\Lambda$-doublet components, which then evolves freely for a time $T$ at the angular frequency $\omega_0$, developing a phase difference of $\delta\,T$ relative to the microwave oscillator, where $\delta = (\omega-\omega_0)$. We choose the free evolution time to be $T=m\lambda/(2v_0)-\tau$, where $v_{0}$ is the mean speed, $\lambda$ is the wavelength in the transmission line, and $m$ is an integer, so that the molecules are now at antinode $m_1+m$. A second microwave pulse of equal duration and amplitude then completes the population transfer with a probability
\begin{align}
P(\delta)&=\frac{4 \pi^2 \sin^2\!\left(\tfrac{X}{4}\right)}{X^4}\times\nonumber\\
& \left[X\!\cos\!\left(\tfrac{X}{4}\right)\!\cos\!\left(\tfrac{\delta T + \beta}{2}\right)\!-\!2\delta  \tau  \sin\!\left(\tfrac{X}{4}\right) \sin\!\left(\tfrac{\delta T + \beta}{2}\right)\right]^2.
\label{Eq:RamseyLineshape}
\end{align}
Here, $X=\sqrt{\pi^{2}+4\delta^{2}\tau^{2}}$ and $\beta$ represents a possible phase shift of the microwave field (modulo $2\pi$) due to the change in position of the molecules between the two pulses. The population in one of the parity states is measured, as a function of $\omega$, by time-resolved laser-induced fluorescence. From the time of flight profile we select molecules with arrival times in the range $t_0 \pm \delta t$, where $t_0$ is the most probable arrival time, and $\delta t \simeq 0.02 t_0$. This ensures that all molecules used in the analysis are sufficiently well localized near the antinodes when the pulses are applied.

\begin{figure*}
\includegraphics[width=\linewidth]{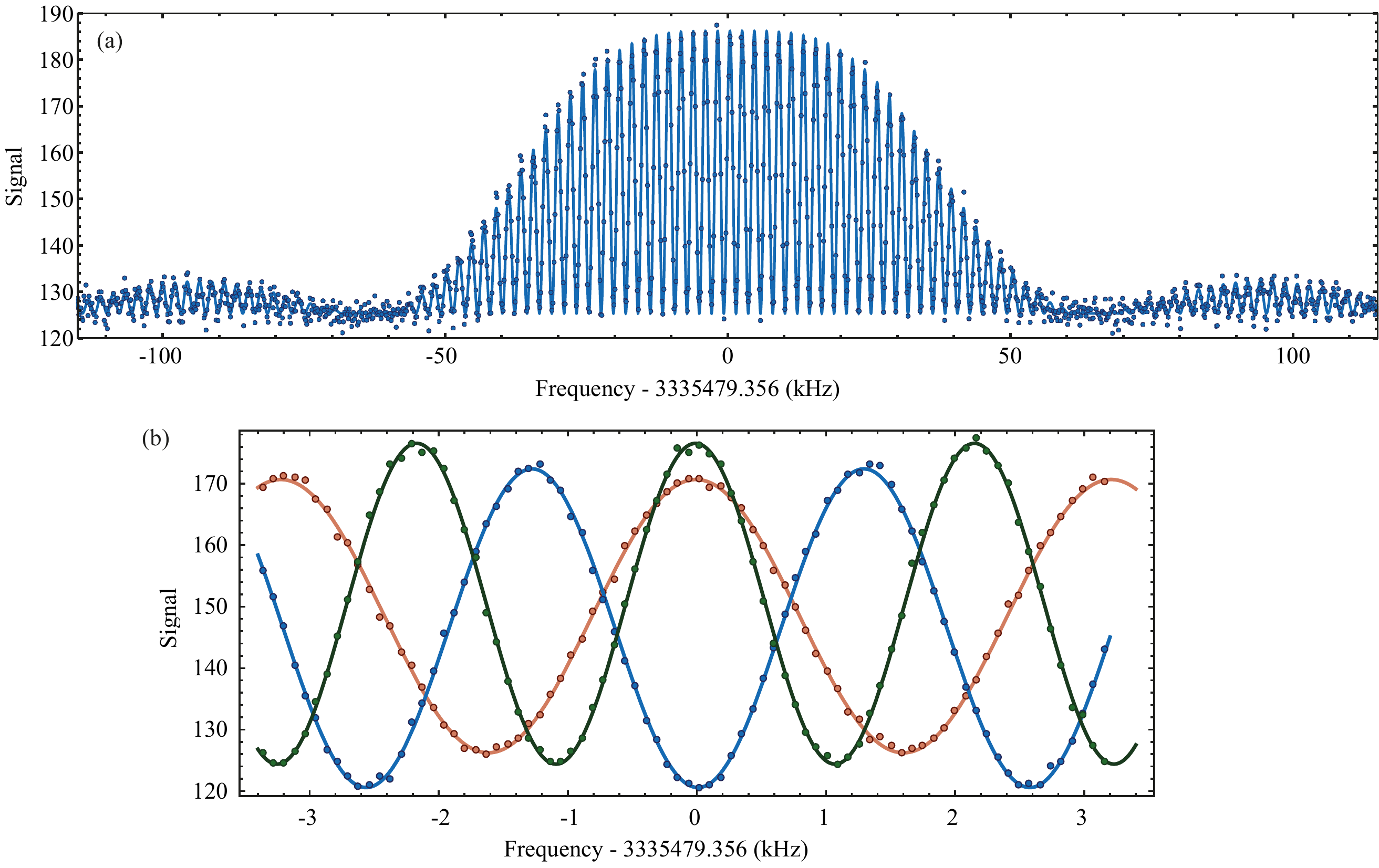}
\caption{\label{Fig:Ramsey} Ramsey interference data. Dots: Measured population in $(1/2^-,1)$ as a function of the microwave frequency using the method of separated oscillating fields. Line: Fit to the data using the model discussed in the text. (a) Wide frequency scan with a free evolution time of $T=428$\,$\mu$s. (b) Narrower frequency scans for three different free evolution times. Green: $458\,\mu$s. Blue: $380\,\mu$s. Red: $302\,\mu$s. }
\end{figure*}

Figure \ref{Fig:Ramsey}(a) shows data measured over a wide frequency range around the $(1/2^+,1)$--$(1/2^-,1)$ resonance. The line is a fit to the model $b + a P(\delta)$, where $b$ and $a$ are constants and $P(\delta)$ is given by equation \,(\ref{Eq:RamseyLineshape}). We set $\beta=0$, and $\tau$ and $T$ to the values used in the experiment, leaving only the offset $b$, the amplitude $a$ (which is negative when $m$ is odd), and the resonance angular frequency $\omega_0$ as fit parameters. The model fits the data well and we use it to determine $\omega_0$. Figure \ref{Fig:Ramsey}(b) shows narrower frequency scans for the same transition and for three different values of $T$ chosen to make $m=4,5$ and 6. Here, as in all our data, we see no dependence of $\omega_0$ on $m$, so we average together measurements taken using different values of $m$. We also see no variation of $\omega_0$ when we subdivide the data according to the arrival time of the molecules.

Our model assumes that $\beta$ is zero, as for a perfect standing wave. Since the standing wave is not perfect, $\beta$ is not exactly zero and there will be a systematic frequency shift of $\delta f = \beta/(2\pi T)$. Let the amplitudes of the two counter-propagating waves be proportional to $(1 \pm \Delta)$, so that $\Delta$ is the amplitude imbalance between them. The microwave electric field is
\begin{equation}
E=A\sqrt{\cos^{2}(k z)+\Delta^{2}\sin^{2}(k z)}\cos(\omega t - \phi),
\end{equation}
where $A$ is an amplitude, $k=2\pi/\lambda$, and (modulo $\pi$)
\begin{equation}
\phi = \tan^{-1}\left(\Delta \tan(k z)\right).
\label{Eq:WavePhase}
\end{equation}
When $\Delta$ is small, $\phi$ changes very slowly around the antinodes. Suppose the first pulse starts when the molecules are at $z_1=z_0 + \delta z_0$, where $z_0$ is the position of an antinode, and the second pulse starts when the molecules are at $z_2 = z_1 + (1+\epsilon)m \lambda/2$. Expanding the trigonometric functions for small displacements around the antinodes (i.e. $\delta z_0 \ll \lambda$, $\epsilon \ll 1$), gives $\delta f \simeq \Delta \epsilon (v_0/\lambda)$, independent of $\delta z_0$ to first order. For the $J=1/2$ measurement, taking $\Delta=0.1$ and $\epsilon=0.05$ (see Methods), we would expect a velocity-dependent frequency shift of $0.06$\,Hz/(m/s).

A second systematic frequency shift is expected due to the motion of the molecules during the two short $\pi/2$ pulses. Including the Doppler shift in the expression for the lineshape in the case of perfect $\pi/2$ pulses, and expanding about the resonance frequency, we find that a travelling wave results in the systematic frequency shift $\delta f_D = (1-4/\pi) (v_0/\lambda)(\tau / T)$. For the counter-propagating waves this is further reduced by the imbalance factor $\Delta$. For the $J=1/2$ measurement, with $\Delta=0.1$, $\tau = 15$\,$\mu$s and $T=450$\,$\mu$s, we expect a shift of $-0.01$\,Hz/(m/s).

We control these, and any other velocity-dependent shifts, by measuring all the frequencies for at least three different values of $v_0$. We find that the measured frequencies shift linearly with $v_0$, and that the gradient, $df/dv_0$, differs for different values of $m_{1}$. The largest gradients we measure are $0.07\pm0.01$\,Hz/(m/s) for the $J=1/2$ measurement, and $0.03\pm0.01$\,Hz/(m/s) for the $J=3/2$ measurement. After extrapolating to zero velocity the results for different values of $m_{1}$ all agree, and so we take the weighted mean of the zero-velocity values as the final frequency. This procedure does not rely on knowledge of $\Delta$.

To minimize systematic shifts due to magnetic fields, the interaction region is magnetically shielded. For a linearly polarized microwave field, and small magnetic field, there is a symmetry in the shifts and amplitudes of the Zeeman sub-components, and therefore no systematic shift. Circular polarization components of the microwave field are strongly suppressed inside the transmission line. For the $J=1/2$ level the spin and orbital angular momentum contributions to the magnetic moment very nearly cancel, and so Zeeman shifts are small. For example, the shift of the $M_F=+1$ component of the $(1/2^+,1)$ level is approximately 0.01\,Hz/nT. For the $J=3/2$ level the Zeeman shifts are far greater, $15.1 M_{F}$\,Hz/nT for $F=1$ and $9.07 M_{F}$\,Hz/nT for $F=2$. To measure the residual field in the interaction region, we applied single $780$\,$\mu$s-long pulses to molecules moving at 570\,m/s, and measured the Zeeman splittings of the $J=3/2$ levels as a function of applied magnetic field in all three directions. Reversal of the applied field measures the residual field in that direction, averaged over the interaction region. We also look for any broadening of the single-pulse lineshape at zero applied field. The residual fields are consistent with zero in all three directions, with upper limits of 3, 56 and 25\,nT along $x$, $y$ and $z$ respectively. To measure the size of magnetic-field related shifts in the Ramsey data, we apply small homogeneous fields in each direction. For the $J=1/2$ transitions, we observe no significant shift at the 1\,Hz level, even for fields as large as 50\,$\mu$T. For the $J=3/2, \Delta F=0$ transitions, the field-induced frequency shifts are all smaller than 0.04\,Hz/nT. Multiplying this by the largest upper limit to the field gives a systematic uncertainty of 2\,Hz. For the $J=3/2,\Delta F=1$ transitions, we could not rule out gradients as large as 0.1\,Hz/nT for $x$, 0.2\,Hz/nT for $y$ and 0.05\,Hz/nT for $z$. Multiplying these by the upper limits to the residual field gives a systematic uncertainty of 11\,Hz for these transitions.

We measured the Stark shift of the $J=1/2$ transition by applying electric fields up to 20\,V/cm between the plates of the transmission line. Measuring the change in this shift upon reversing the applied field sets an upper limit to uncontrolled Stark shifts of 0.1\,Hz. We checked for systematic shifts depending on microwave power or probe laser detuning, but these were negligible. Frequency shifts due to collisions in the beam, spurious frequency sidebands, black-body radiation \cite{Vanhaecke(1)08}, the motional Stark shift, and the second-order Doppler shift, are all negligible at the current accuracy level of $10^{-9}$.

Table \ref{Tab:Frequencies} shows our measured frequencies along with their uncertainties, which are obtained by adding the statistical and systematic uncertainties in quadrature. The systematic uncertainty is negligible for all but the last two, where it is 11\,Hz. The three $J=1/2$ measurements are consistent with the previous best laboratory \cite{McCarthy(1)06} and astrophysical \cite{Rydbeck(1)74, Zuckerman(1)75} measurements and are 300 times more precise than these. We know of no previous laboratory measurements of the $J=3/2$ frequencies. There are astrophysical measurements of the first two $J=3/2$ transitions in the table \cite{Ziurys(1)85}, and our results are consistent with these and about 300 times more precise.

\begin{table}
\begin{center}
\begin{tabular}{l|l}
\hline
Transition & Frequency (Hz)\\
\hline
$(1/2^+,1)$ -- $(1/2^-,1)$ & $3335479356 \pm 3$ \\
$(1/2^+,0)$ -- $(1/2^-,1)$ & $3349192556 \pm 3$ \\
$(1/2^+,1)$ -- $(1/2^-,0)$ & $3263793447 \pm 3$ \\
$(3/2^+,2)$ -- $(3/2^-,2)$ & $701677682 \pm 6$ \\
$(3/2^+,1)$ -- $(3/2^-,1)$ & $724788315 \pm 16$ \\
$(3/2^+,1)$ -- $(3/2^-,2)$ & $703978340 \pm 21$ \\
$(3/2^+,2)$ -- $(3/2^-,1)$ & $722487624 \pm 16$ \\
\hline
\end{tabular}
\end{center}
\caption{
\label{Tab:Frequencies}
Measured $\Lambda$-doublet frequencies, with $1\sigma$ uncertainties. Levels are labelled with the notation $(J^p,F)$.
}
\end{table}

\textbf{Constraining changes of $\alpha$ and $\mu$ in the Milky Way.} With these new measurements, and existing astronomical data, we can set useful bounds on possible variations of $\alpha$ and $\mu$ with matter density. We study five interstellar sources where the density is typically $10^{19}$ times smaller than on Earth. It is necessary to compare two spectral lines from each source, so that the Doppler shift can be removed, and for high sensitivity to changes in the constants these two should have very different sensitivity coefficients. In one of the sources the $J=1/2$ and $J=3/2$ $\Lambda$-doublet transitions of CH are observed together, and in the other four the $J=1/2$ CH transition is observed together with the ground state $\Lambda$-doublet transition of OH. Astronomical spectra are plotted on a velocity scale which is determined from the Doppler shift of the received radiation. For each selected source the spectral lines are narrow and have similar shapes, and we assume that the molecules producing them have the same velocities. To obtain our bounds, we assume that there is either a change in $\alpha$ or a change in $\mu$, but not both (this is a limitation of our analysis). Then, the fractional change in $\alpha$ is
\begin{equation}
\frac{\Delta\alpha}{\alpha} = \frac{1}{(K_{\alpha}^{(2)} - K_{\alpha}^{(1)})c}\left[\Delta v_{12} + c\left(\frac{\omega_{\text{lab}}^{(1)}}{\omega_{\text{astro}}^{(1)}} - \frac{\omega_{\text{lab}}^{(2)}}{\omega_{\text{astro}}^{(2)}}\right)\right],
\label{Eq:dalpha}
\end{equation}
with a similar equation for the change in $\mu$. Here, $\Delta v_{12}=v^{(1)}-v^{(2)}$ is the difference between the measured velocities of lines 1 and 2 and is determined from the published spectra, $K_{\alpha}^{(i)}$ is the sensitivity coefficient for transition $i$, $\omega_{\text{lab}}^{(i)}$ is the frequency measured in the laboratory, and $\omega_{\text{astro}}^{(i)}$ is the frequency used to define the velocity scale in the astronomical measurement. This is usually the accepted value for the transition frequency at the time of the measurement, is specified along with the published spectra, and carries no uncertainty. The laboratory measurements presented here allow the second term in square brackets to be determined to high accuracy, and so the uncertainty in $\Delta\alpha/\alpha$ is now due entirely to the uncertainty in $\Delta v_{12}$.

Table \ref{Tab:Results} shows the results obtained from the five selected interstellar sources. We find the central velocities by fitting Gaussian profiles to the spectral lines. Where there are multiple velocity components, we fit multiple Gaussians. In each source there are multiple velocity components and/or multiple hyperfine components, and so we obtain several different values for the velocity difference. Their standard deviation is typically several times the individual uncertainties of the fits, probably because the underlying velocity distributions are not Gaussian. For each source, we use the mean and standard deviation of the individual results to give a central value and uncertainty for $\Delta v_{12}$. The angular size of the observed part of the source (the beam size) is often different for the two species. This can result in systematic velocity differences. This is not significant for CasA or W51 where the sources are smaller or comparable in size to the smallest beam size used. Neither is it significant for Heiles cloud 2 and L134N, because the beam size used for the CH and OH observations were nearly the same, and maps of these clouds show that velocity variation is less than 0.2\,km/s over a much wider angular scale than this beam size \cite{Rydbeck(1)74}. For the observations of RCW 36, which is an extended source, the beam size for OH was twice that for CH. However, maps of this cloud \cite{Whiteoak(1)77} show that the observed region is centred on an intense localized clump whose size is approximately the OH beam size, and that any systematic velocity differences due to different beam sizes will be smaller than our quoted uncertainty.

Using $\Delta v_{12}$, the measured values of $\omega_{\text{lab}}^{(i)}$ for CH and OH \cite{Hudson(1)06}, and the quoted values of $\omega_{\text{astro}}^{(i)}$, we calculate the quantity in square brackets in equation \,(\ref{Eq:dalpha}), which we label as $\Delta v_{12}'$ in Tab.\,\ref{Tab:Results}. From these, and the calculated sensitivity coefficients for OH \cite{Kozlov(1)09} and CH \cite{Nijs(1)12}, we find values for $\Delta\alpha/\alpha$ and $\Delta\mu/\mu$. Taking the weighted mean of their values, we arrive at the results $\Delta\alpha/\alpha = (0.3 \pm 1.1) \times 10^{-7}$ and $\Delta\mu/\mu = (-0.7 \pm 2.2) \times 10^{-7}$, giving us $1\sigma$ upper bounds of $|\Delta \alpha/\alpha| < 1.4 \times 10^{-7}$ and $|\Delta \mu/\mu| < 2.9 \times 10^{-7}$.

\begin{table*}
\footnotesize
\begin{center}
\begin{tabular}{c|c|c|c|c|c|c|c|c}
\hline
Source & Transition 1 & Transition 2 & $v$(km/s) & $\Delta v_{12}$(km/s) & $\Delta v_{12}'$(km/s) & $\tfrac{\Delta \alpha}{\alpha}(10^{-7})$ & $\tfrac{\Delta\mu}{\mu}(10^{-7})$ & Ref \\
 \hline
 G111.7-2.1(CasA) & CH(3264,3335,3349) & OH(1667) & -1.4, 0 & $-0.01(0.09)$ & $-0.08(0.11)$ & $1.5(2.0)$ & $-3.1(4.1)$ & \cite{Rydbeck(1)74, Rydbeck(1)76} \\
 G265.1+1.5(RCW36) & CH(3264,3335) & OH(1612,1665,1667,1721) & 6.8 & $0.06(0.19)$ & $0.04(0.16)$ & $0.9(3.1)$ & $1.9(6.4)$ & \cite{Whiteoak(1)78} \\
 G174.3-13.4(Heiles2) & CH(3264,3335,3349) & OH(1612,1665,1667,1721) & 5.8 & $0.00(0.19)$ & $-0.02(0.19)$ & $0.6(3.6)$ & $-1.2(7.4)$ & \cite{Rydbeck(1)76, Turner(1)73}\\
 G6.0+36.7(L134N) & CH(3264,3335,3349) & OH(1665,1667) & 2.5 & $0.05(0.13)$ & $-0.12(0.13)$ & $2.3(2.4)$ & $-4.8(5.0)$ & \cite{Rydbeck(1)76, Turner(1)73}\\
 G49.5-0.4(W51) & CH(702) & CH(3264,3335,3349) & 65 & $-0.85(0.53)$ & $-0.48(0.55)$ & $-1.8(2.0)$ & $3.6(4.1)$ & \cite{Ziurys(1)85} \\
 \hline
\end{tabular}
\end{center}
\caption{ Analysis of astronomical data. The columns are: the direction of the source, the transitions used (labelled by the frequencies of the hyperfine components in MHz), the velocity component(s) used, the measured velocity difference, the velocity difference corrected for any differences between the nominal and laboratory frequencies, the derived values of $\Delta \alpha / \alpha$ and $\Delta \mu / \mu$, the references where the spectra are given. $1\sigma$ uncertainties are given in parentheses.
\label{Tab:Results}
}
\end{table*}

\textbf{Outlook.} Our analysis demonstrates the high sensitivity of this method. With dedicated astronomical measurements, far higher precision could be reached and the study could be extended to sources at high red-shift. The scope is particularly great for studying sources where the 3.3 and 0.7\,GHz transitions are observed simultaneously. There is currently a shortage of such data, and the comparison provides excellent sensitivity to variation of the constants and is relatively immune to systematic shifts that can arise when analyzing transitions in different species. The relatively low intensity of the CH lines requires that large telescopes are used, but there are several that are suitable. The 100\,m telescope of the Max-Planck-Gesellschaft covers the entire frequency range between the two $\Lambda$-doublets, while the 100\,m Green Bank Telescope covers the frequency range from 0.29 to 2.6 GHz and so is suitable for measuring both transitions at moderate red-shifts. For higher collecting surface, the 300\,m telescope at Arecibo can be used. It covers frequencies from 3.3\,GHz down to 1.225\,GHz and selected frequency bands between 690 and 312\,MHz, and so could be used to study both transitions over a wide range of red-shifts. In a few years, even higher sensitivity will be provided by a new 500\,m telescope being built in the Guizhou province of China. To circumvent problems of rf interference, interferometric measurements are efficient. The Very Large Array provides such measurements with high sensitivity and currently covers all frequencies between 1 and 3.3\,GHz. Looking further ahead, the Square Kilometer Array, to be constructed in South Africa and Australia, will provide unprecedented sensitivity. Being an interferometric facility located in areas of low interference, this telescope should be able to measure the two $\Lambda$-doublets of CH in a large number of interstellar clouds, both at low and at high redshift. With a reasonable precision of 0.1\,km/s in the measurement of $\Delta v_{12}$, and a modest dataset, it will be possible to reach below the $10^{-8}$ level for both $\Delta \alpha/\alpha$ and $\Delta \mu/\mu$. The laboratory frequencies are now so accurate that their uncertainties are unlikely to be a limiting factor in any future measurements.

\textbf{Methods.} The setup for measuring the frequencies is shown in Figure \ref{Fig:Setup}. A pulsed supersonic beam of CH radicals, with a repetition rate of 10\,Hz, is produced by photo-dissociation of CHBr$_3$ (96\% purity stabilized in ethanol) in a 4\,bar carrier gas of either He, Ne, Ar or Kr \cite{Lindner(1)98, Romanzin(1)06}. The measured mean speeds of the CH for each of these carrier gases are $v_0=1710$, 810, 570, and 420\,m/s, and the translational temperature is typically 0.4--0.5\,K. About 95\% of the molecules are formed in the $J=1/2$ state. The mixture, formed by bubbling the carrier gas through liquid bromoform, expands through the 1\,mm-diameter nozzle of a pulsed solenoid valve into the vacuum chamber. The light from the photo-dissociation laser (220\,mJ, 20\,ns, 248\,nm) is focused in front of the nozzle to a spot 1\,mm high (along $y$) and $4$\,mm wide (along $z$).

At $z\!=\!86$\,mm the molecules pass through a 2\,mm diameter skimmer and enter the second chamber, where the pressure is below $10^{-7}$\,mbar. They pass through the state selector and transmission line resonator described below, and are then detected at $z=780$\,mm by driving the $\text{A}^2\Delta (v=0) \leftarrow \text{X}^2\Pi (v=0)$ transition with a laser, and imaging the resulting fluorescence onto a photomultiplier tube. In this way, the time-of-flight profile of each molecular pulse is recorded with a temporal resolution of about 5\,$\mu$s. The probe laser beam comes from a frequency-doubled continuous-wave titanium-sapphire laser. The beam propagates along $x$, is linearly polarized along $z$, has a wavelength near 430.15\,nm, and has a power of 5\,mW in a rectangular-shaped beam, 4\,mm high and $1.4$\,mm wide. We measure the population in the $(1/2^-,F)$ levels for the $J=1/2$ measurement, and in the $(3/2^+,F)$ levels for the $J=3/2$ measurement, by driving the transitions to the $J=3/2$ and $J=5/2$ levels of $\text{A}^2\Delta (v=0,N=2)$. The frequencies of these transitions, which are labelled R$_{\text{22ff}}(1/2)$ and R$_{\text{11ff}}(3/2)$, are given in reference \cite{Zachwieja(1)95}. The CH density at the detector is approximately $10^{6}$\,cm$^{-3}$.

The state selector is situated at $z=241$\,mm. For the $J=1/2$ measurement it is a laser beam of the same frequency, polarization and propagation direction as the detection laser. This beam depletes the population of the negative parity component. For the $J=3/2$ measurement, we drive population from $(1/2^-,F)$ to the $(3/2^+,F')$ state, using approximately 10$\mu$W of radiation near 533\,GHz. This mm-wave radiation is the amplified 54th harmonic of a frequency synthesizer. The radiation is collimated by a teflon lens of 30\,mm focal length, giving a 10\,mm diameter beam which propagates along $x$ and is polarized along $y$. The efficiency of the population transfer is about 50\%.

The microwave transmission line resonator is built from a pair of 30\,mm wide parallel copper plates separated by 5\,mm along $y$. At the $\Lambda$-doublet frequencies of interest, the transmission line can only support a TEM$_{00}$ mode polarized along $y$. Microwave radiation is launched into one end from a semi-rigid coaxial cable. The other end is not terminated and so a resonator with a quality factor of approximately 10 is formed. For each frequency component measured, the length was cut so that the transmission line had a resonance at that frequency. The length was approximately 480\,mm for the $J=1/2$ measurement and 440\,mm for the $J=3/2$ measurement. The microwave radiation is generated by a synthesizer phase locked to a GPS frequency reference whose fractional accuracy is better than $10^{-13}$. A fast, high isolation switch controls the microwave pulse duration. The transmission line is housed inside a mu-metal magnetic shield. For the $J=1/2$ measurement a single shield is used, while for the $J=3/2$ measurement we use a pair of nested shields. The outer shield extended into the region of the state selector, and a pair of holes cut into this shield allowed the mm-wave beam to pass through. Coils inside the shields allow us to apply well-controlled magnetic fields.

For each transition, we first make an approximate measurement of the transition frequency by applying a single microwave pulse to the molecules and measuring the population as a function of the frequency. Due to the two counter-propagating waves we observe two peaks separated by twice the Doppler shift. Their relative amplitudes indicates that $\Delta\approx 0.1$. The mean frequency of these peaks is the transition frequency. With the frequency fixed at this value, we map out the standing microwave field by driving the transition with a 15\,$\mu$s pulse and scanning the time when it is applied. The times at which the resonance is largest are the times when the molecules pass each of the antinodes. Next, we apply a short pulse when the molecules are at an antinode, and we scan the microwave power. We observe Rabi oscillations in the population, and thus determine the power needed to drive a $\pi$-pulse. We do this for each of the antinodes. With the parameters now fixed, we measure the transition frequency using the Ramsey method of separated oscillating fields \cite{Ramsey(1)49}, as described in the main text. Compared to cw excitation, this method has higher resolution and is relatively immune to broadening and distortion of the lineshape caused by non-uniform fields.

\vspace{1cm}

\textbf{Acknowledgements.} We thank Ben Sauer and Jony Hudson for their helpful advice. This work was supported in the UK by the EPSRC and the Royal Society.

\vspace{0.5cm}

\textbf{Author contributions.}
The experiment was designed and built by S.T., S.K.T, R.J.H. and M.R.T. The data were collected by S.T., S.K.T., R.J.H. and H.J.L. The laboratory data were analyzed by S.T., R.J.H. and M.R.T. The astronomical data were analyzed by M.R.T. The manuscript was drafted by S.T.,  M.R.T. and E.A.H and the outlook prepared by C.H. The work was inspired by M.G.K. and the team led by M.R.T and E.A.H. All authors discussed the methodology and the results, and improved the manuscript.

\vspace{0.5cm}

\textbf{Competing financial interests.}
The authors declare that they have no competing financial interests.


\begin{thebibliography}{20}

\bibitem{Uzan(1)03} Uzan, J. P. The fundamental constants and their variation: observational and theoretical status. \textit{Rev. Mod. Phys.} \textbf{75}, 403 (2003).

\bibitem{Uzan(1)11} Uzan, J. P. Varying Constants, Gravitation and Cosmology. \textit{Living Rev. Relativ.} \textbf{14}, 2 (2011).

\bibitem{Khoury(1)04} Khoury, J. \& Weltman, A. Chameleon Fields: Awaiting Surprises for Tests of Gravity in Space. \textit{Phys. Rev. Lett.} \textbf{93}, 171104 (2004).

\bibitem{Brax(1)04} Brax, P., van de Bruck, C., Davis, A.-C., Khoury, J. \& Weltman, A. Detecting dark energy in orbit: The cosmological chameleon. \textit{Phys. Rev. D} \textbf{70}, 123518 (2004).

\bibitem{Olive(1)08} Olive, K. A. \& Pospelov, M. Environmental dependence of masses and coupling constants. \textit{Phys. Rev. D} \textbf{77}, 043524 (2008).

\bibitem{Rosenband(1)08} Rosenband, T. {\it et al.}, Frequency ratio of Al+ and Hg+ single-ion optical clocks; metrology at the 17th decimal place. \textit{Science} \textbf{319}, 1808 (2008).

\bibitem{Blatt(1)08} Blatt, S. {\it et al.} New limits on coupling of fundamental constants to gravity using $^{87}$Sr optical lattice clocks. \textit{Phys. Rev. Lett.} \textbf{100}, 140801 (2008).

\bibitem{Webb(1)11} Webb, J. K. {\it et al.} Indications of a spatial variation of the fine structure constant. \textit{Phys. Rev. Lett.} \textbf{107}, 1 (2011).

\bibitem{Molaro(1)08} Reimers, D., Agafonova, I. I., \& Levshakov, S. A. Bounds on the fine structure constant variability from Fe II absorption lines in QSO spectra. \textit{Eur. Phys. J. Spec. Top.} \textbf{163}, 173 (2008).

\bibitem{Agafonova(1)11} Agafonova, I. I., Molaro, P., Levshakov, S. A. \& Hou, J. L. First measurement of Mg isotope abundances at high redshifts and accurate estimate of $\Delta\alpha/\alpha$. \textit{Astron. Astrophys.} \textbf{529}, A28 (2011).

\bibitem{Levshakov(1)12} Levshakov, S. A. {\it et al.} An upper limit to the variation in the fundamental constants at redshift z=5.2. \textit{Astron. Astrophys.} \textbf{540}, L9 (2012).

\bibitem{Weiss(1)12} Wei\ss, A. {\it et al.} On the variations of fundamental constants and active galactic nucleus feedback in the quasi-stellar object host galaxy RXJ0911.4+0551 at z=2.79. \textit{Astrophys. J.} \textbf{753}, 102 (2012).

\bibitem{Kanekar(1)10} Kanekar, N., Chengalur, J. N. \& Ghosh, T. Probing fundamental constant evolution with redshifted conjugate-satellite OH lines. \textit{Astrophys. J. Lett.} \textbf{716}, L23 (2010).

\bibitem{Rahmani(1)12} Rahmani, H. {\it et al.} Constraining the variation of fundamental constants at z~1.3 using 21-cm absorbers. \textit{Mon. Not. R. Astron. Soc.} \textbf{425}, 556 (2012).

\bibitem{Bagdonaite(1)13} Bagdonaite, J. {\it et al.} A stringent limit on a drifitng proton-to-electron mass ratio from alcohol in the early universe. \textit{Science} \textbf{339}, 46 (2013).

\bibitem{Levshakov(3)10} Levshakov, S. A., Molaro, P. \& Reimers, D. Searching for spatial variations of $\alpha^{2}/\mu$ in the Milky Way. \textit{Astron. Astrophys.} \textbf{516}, A113 (2010).

\bibitem{Levshakov(1)11} Levshakov, S. A., Kozlov, M. G. \& Reimers, D. Methanol as a tracer of fundamental constants. \textit{Astrophys. J.} \textbf{738}, 26 (2011).

\bibitem{Levshakov(1)10} Levshakov, S. A. {\it et al.} Searching for chameleon-like scalar fields with the ammonia method. \textit{Astron. Astrophys.} \textbf{512}, A44 (2010).

\bibitem{Levshakov(2)10} Levshakov, S. A. {\it et al.} Searching for chameleon-like scalar fields with the ammonia method. II. Mapping of cold molecular cores in NH$_3$ and HC$_3$N lines. \textit{Astron. Astrophys.} \textbf{524}, A32 (2010).

\bibitem{Kozlov(1)09} Kozlov, M. G. $\Lambda$-doublet spectra of diatomic radicals and their dependence on fundamental constants. \textit{Phys. Rev. A} \textbf{80}, 022118 (2009).


\bibitem{Nijs(1)12} de Nijs, A. J., Ubachs W. \& Bethlem, H. L. Sensitivity of rotational transitions in CH and CD to a possible variation of fundamental constants. \textit{Phys. Rev. A} \textbf{86}, 032501 (2012).

\bibitem{Shelkovnikov(1)08} Shelkovnikov A., Butcher, R. J., Chardonne, C. \& Amy-Klein, A. Stability of the proton-to-electron mass ratio. \textit{Phys. Rev. Lett.} \textbf{100}, 150801 (2008).

\bibitem{Bethlem(1)08} Bethlem, H. L., Kajita, M., Sartakov, B., Meijer, G. \& Ubachs, W. Prospects for precision measurements on ammonia molecules in a fountain. \textit{Eur. Phys. J. Special Topics} \textbf{163}, 55 (2008).

\bibitem{Bethlem(1)09} Bethlem, H. L. \& Ubachs, W. Testing the time-invariance of fundamental constants using microwave spectroscopy on cold diatomic radicals. \textit{Faraday Discuss.} \textbf{142}, 25 (2009).

\bibitem{Dickenson(1)13} Dickenson, G. D. {\it et al.} Fundamental vibration of molecular hydrogen. \textit{Phys. Rev. Lett.} \textbf{110}, 193601 (2013).

\bibitem{Salumbides(1)13} Salumbides, E. J. Bounds on fifth forces from precision measurements on molecules. \textit{Phys. Rev. D} \textbf{87}, 112008 (2013).

\bibitem{Rydbeck(1)73} Rydbeck, O. E. H., Elld\'er, J. \& Irvine, W. M. Radio detection of interstellar CH. \textit{Nature} \textbf{246}, 466 (1973).

\bibitem{Turner(1)74} Turner, B. E. \& Zuckerman, B. Microwave detection of interstellar CH. \textit{Astrophys. J.} \textbf{187}, L59 (1974).

\bibitem{Robinson(1)74} Robinson, B. J., Gardner, F. F., Sinclair, M. W. \& Whiteoak, J. B. Absorption and emission by interstellar CH at 9\,cm. \textit{Nature} \textbf{248}, 31 (1974).

\bibitem{Rydbeck(1)74} Rydbeck, O. E. H., Elld\'er, J., Irvine, W. M., Sume, A. \& Hjalmarson, \AA. Radio astronomical determination of ground state transition frequencies of CH. \textit{Astron. Astrophys.} \textbf{34}, 479 (1974).

\bibitem{Zuckerman(1)75} Zuckerman, B. \& Turner, B. E. Observations of interstellar CH and a study of its chemistry and excitation. \textit{Astrophys. J.} \textbf{197}, 123 (1975).

\bibitem{Rydbeck(1)76} Rydbeck, O. E. H. {\it et al.} Radio observations of interstellar CH. I. \textit{Astrophys. J. Suppl. S.} \textbf{31}, 333 (1976).

\bibitem{Hjalmarson(1)77} Hjalmarson, \AA. {\it et al.} Radio observations of interstellar CH. II. \textit{Astrophys. J. Suppl. S.} \textbf{35}, 263 (1977).

\bibitem{Whiteoak(1)78} Whiteoak, J. B., Gardner, F. F. \& Sinclair, M. W. Observations of the 3.3-GHz ground-state transitions of interstellar CH. \textit{Mon. Not. R. Astron. Soc.} \textbf{184}, 235 (1978).

\bibitem{Genzel(1)79} Genzel, R., Downes, D., Pauls, T., Wilson T. L. \& Bieging, J. Interstellar CH: Excitation temperatures and abundance relative to H$_2$CO. \textit{Astron. Astrophys.} \textbf{73}, 253 (1979).

\bibitem{Ziurys(1)85} Ziurys, L. M. \& Turner, B. E., Detection of interstellar rotationally excited CH. \textit{Astrophys. J.} \textbf{292}, L25 (1985).

\bibitem{Whiteoak(1)80} Whiteoak, J. B., Gardner, F. F. \& H\"oglund, B. The detection of CH in external galaxies. \textit{Mon. Not. R. Astron. Soc.} \textbf{190}, 17 (1980)

\bibitem{Nieves(1)13} Nieves, Y., Salter, C. J., Minchin, R. F. \& Ghosh, T., The interstellar media of the luminous infra-red Galaxies, IC860 and Zw049.057. \textit{American Astronomical Society, AAS Meeting \#221, \#349.25} (2013).


\bibitem{Ramsey(1)49} Ramsey, N. F. A new molecular beam resonance method. \textit{Phys. Rev.} \textbf{76}, 996 (1949);
\bibitem{Ramsey(1)50} Ramsey, N. F. A molecular beam resonance method with separated oscillating fields. \textit{Phys. Rev.} \textbf{78}, 695 (1950).

\bibitem{Vanhaecke(1)08} Vanhaecke, N. \& Dulieu, O. Precision measurements with polar molecules: the role of the black body radiation. \textit{Mol. Phys.} \textbf{105}, 11 (2008).

\bibitem{McCarthy(1)06} McCarthy, M. C., Mohamed, S., Brown, J. M. \& Thaddeus, P. Detection of low-frequency lambda-doublet transitions of the free $^{12}$CH and $^{13}$CH radicals. \textit{Proc. Natl. Acad. Sci.} \textbf{103}, 12263 (2006).

\bibitem{Turner(1)73} Turner, B. E. Nonthermal OH emission in interstellar dust clouds II. \textit{Astrophys. J} \textbf{186}, 357 (1973).


\bibitem{Whiteoak(1)77} Whiteoak, J. B. and Gardner F. F., H$_{2}$CO and OH observations of a molecular cloud near RCW 36. \textit{P. Astron. Soc. Aust.} \textbf{3}, 147 (1977).

\bibitem{Hudson(1)06} Hudson, E. R., Lewandowski, H. J., Sawyer, B. C. \& Ye, J. Cold molecule spectroscopy for constraining the evolution of the fine strucutre constant, \textit{Phys. Rev. Lett.} \textbf{96}, 143004 (2006).

\bibitem{Lindner(1)98} Lindner, J., Ermisch, K. \& Wilhelm, R. Multi-photon dissociation of CHBr$_3$ at 248 and 193\,nm: observation of the electronically excited CH(A$^{2}\Delta$) product. \textit{Chem. Phys.} \textbf{238}, 329 (1998).

\bibitem{Romanzin(1)06} Romanzin, C. {\it et al.} CH radical production from 248 nm photolysis or discharge-jet dissociation of CHBr$_3$ probed by cavity ring-down absorption spectroscopy.  \textit{J. Chem. Phys.} \textbf{125}, 114312 (2006).

\bibitem{Zachwieja(1)95} Zachwieja, M., New investigations of the A$^{2}\Delta$-X$^{2}\Pi$ band system in the CH radical and a new reduction of the vibration-rotation spectrum of CH from the ATMOS spectra. \textit{J. Mol. Spectrosc.} \textbf{170}, 285 (1995).

\end{thebibliography}
\end{document}